\documentclass[10pt]{article}
\usepackage[OE]{express}

\pdfoutput=1
\usepackage{graphicx}
\usepackage{caption}
\usepackage{subcaption}
\usepackage{dcolumn}
\usepackage{bm}
\usepackage{color} 

\newcolumntype{P}[1]{>{\centering\arraybackslash}p{#1}}
\newcolumntype{M}[1]{>{\centering\arraybackslash}m{#1}}
\usepackage{multirow}
\usepackage{array,booktabs}
\usepackage{ctable}

\begin{document}


\title{Comparative study of afterpulsing behavior and models in single photon counting avalanche photo diode detectors
}

\author{Abdul Waris Ziarkash,\authormark{1, \textdagger} Siddarth Koduru Joshi,\authormark{1,*,\textdagger} Mario Stip\v{c}evi\'c,\authormark{2} and Rupert Ursin\authormark{3,**}}

\address{\authormark{1}Institute for Quantum Optics and Quantum Information, Austrian Academy of Sciences Vienna A-1090\\

\authormark{2}Ruder Boskovic Institute, Center of Excellence for Advanced Materials and Sensors and Division of Experimental Physics, Zagreb 10000, Croatia\\

\authormark{3}Institute for Quantum Optics and Quantum Information, Austrian Academy of Sciences, Vienna A-1090\\ Vienna Center for Quantum Science and Technology (VCQ).}
\authormark{\textdagger} These authors contributed equally and are joint first authors.\\
\email{\authormark{*}Siddarth.Koduru.Joshi@oeaw.ac.at.} 
\email{\authormark{**}Rupert.Ursin@oeaw.ac.at.} 

\date{\today}

%


\begin{abstract}
Single-photon avalanche diode (SPAD) detectors, have a great importance in fields like quantum key distribution, laser ranging, florescence microscopy, etc. Afterpulsing is a non-ideal behavior of SPADs that adversely affects any application that measures the number or timing of detection events. Several studies based on a few individual detectors, derived distinct mathematical models from semiconductor physics perspectives. With a consistent testing procedure and statistically large data sets, we show that different individual detectors - even if identical in type, make, brand, etc. - behave according to fundamentally different mathematical models. Thus, every detector must be characterized individually and it is wrong to draw universal conclusions about the physical meaning behind these models. We also report the presence of high-order afterpulses that are not accounted for in any of the standard models.
\end{abstract}



\section{Introduction}
Applications of single-photon detectors, which focus on the timing of a very weak optical signal, mostly use single-photon avalanche diodes (SPADs) operated above their breakdown voltage in Geiger mode. Such as in photonics, quantum processing tasks, laser ranging, fluorescence microscopy, neural imaging with blood flow tomography, contrast-enhanced MRI, two-photon luminescence imaging, astronomical telescopes, etc.~\cite{eisaman2011invited, hadfield2009single, denk1990two, kuhl1982quantifying, wagner2003contrast, wang2005vitro, holder2006first}. They are widely manufactured and sold by many different companies and as such exhibit different properties/behavior. To efficiently and accurately perform these experiments we must account for all non-ideal behavior of the detectors used~\cite{Brown86,DaSilva2011}. This is especially true for sensitive applications like quantum communication, because the security of any real world implementation (i.e. with a high transmission loss) depends on the devices used. Hence the precise modeling/characterization of the non-ideal behavior of single photon detectors (and all other components of the quantum communication device) are critical for practical security proofs.

An ideal single-photon detector generates one and only one electric pulse for every incident photon. 
However, in a real detector, it is possible that a single incident photon results in more than one electrical pulse per incident photon. This is known as afterpulsing. \cite{Brown86} 
In this work we, approach this problem from the perspective of an end user. Consequently, we use an ipso-facto definition of an ``afterpulse'' as any pulse in addition to and following an isolated detection event and its subsequent dead time, regardless of its etiology. This behavior has been extensively studied due to its importance in semiconductor physics in general. It has been suggested that afterpulsing can be linked to charges trapped in the deep levels of the semiconductor's band structure and released at a later time~\cite{Cova1991} as well as to an other causes (such as those described in Refs.~\cite{Stip2009,Polyakov2007,Stipcevic2014,Ware2006} 
  
Afterpulsing has different implications, depending on the application of the detector used. It can result in an overestimation of the total count rate by up to 10\,\% as well as a reduction of the duty cycle and detection efficiency due to the increased dead times. In fluorescence microscopy it could lead to an overestimation of the concentration of fluorophores. In quantum communication the overestimation of coincidence events leads to a larger Quantum Bit Error Rate (QBER). Afterpulsing can also adversely affect security~\cite{Jain2014}. Afterpulsing also poses  a significant problem to the measurement of photon arrival times both in quantum communication protocols as well as for laser ranging. Hence a proper characterization of the afterpulse behaviour could improve the accuracy of several measurements in metrology in general. 

Several previous works have attempted to characterize the afterpulse behavior and fit the results to various models~\cite{Giudice2003,Cova1991,Itzler2012,Humer2015,Horoshko2015,Jensen2006,Stip2013}. Worryingly, these papers do not agree on the most suitable model describing the statistics of the arrival times. Most of these studies have focused on one make/manufacture of detector at a time. {In this work we take a more comprehensive approach and perform a comparative study on 3 different makes of detectors (for each make we compare up to 12 individual detectors with almost consecutive serial numbers). We test the universality of the various theoretical models and attempt to simply resolve the conflicts raised by previous works.} To do so we eschew the common practice of studying the timing auto-correlation of the detector signals in favor of the more comprehensive timing cross-correlation between the emission of a photon and all the following detector signals. We describe this procedure in Section~\ref{sec:setup}. We then fit to the various standard models described in Section~\ref{sec:models} and compare these fits in Section ~\ref{sec:fitting results}. Due to our use of cross-correlations we were also able to detect higher orders of afterpulses which we present in Section~\ref{sec:higherorder} and finally in Section~\ref{sec:noise} we discuss the corrections that can be applied.

\section{Canonical models of afterpulsing}\label{sec:models}
Much of the previous work on afterpulsing in SPADs focused on modeling their behavior. In this section we present the three most common models. We go on to show the inadequacy of all of these models in Section~\ref{sec:fitting results}. 
The characteristic decay of the afterpulse probability was sometimes thought  to depend on the deep level in which the charge is trapped.
Initial models~\cite{Giudice2003,Cova1991,Yen2008,Humer2015,Korzh2015} considered the decay from several distinct deep levels and proposed the ``multiple exponential model'' where the afterpulsing probability ($P_{exp} (t)$) at a time t is given by:
\begin{equation}\label{eq:multiexp}
	P_{exp}(t) = \sum_{k} A_k e^{-\frac{t}{\tau_{k}}} + d,
\end{equation} 
where $\tau_k$ and $A_k$ is the de-trapping lifetime and amplitude factor for the $k^{\mathrm{th}}$ deep level, and d is the offset due to noise counts.

Furthermore \cite{Itzler2012,Itzler2011} considered a continuum of deep levels, in InGaAs/InP detectors, that could trap a charge and found that the power law model was a good empirical fit. Here the afterpulse probability ($P_{pow} (t)$) is given by: 

\begin{equation}\label{eq:powerlaw}
	P_{pow}(t) = A \cdot t^{- \lambda} + d,
\end{equation}
where $\lambda$ is a effective decay constant and $A$ is the initial afterpulse probability.

In an another attempt to create a more physically meaningful model Ref.~\cite{Horoshko2015} derived the ``hyperbolic Sinc model'' from the Arrhenius law (once more assuming a continuum of levels), where the afterpulse probability ($P_{sinc}(t)$) is given by:
\begin{equation}\label{eq:sinhclaw}
	P_{sinc}(t) = 2 \cdot A \cdot \frac{ \text{sinh}(\Delta \cdot t)}{t} \cdot e^{- \gamma \cdot t} + d,
\end{equation}
where $\Delta$ and $\gamma$ are both functions of the minimum and maximum energies of the deep levels in which charges may be trapped.

In this paper, we compare the above three standard mathematical models (see Equations~\ref{eq:multiexp}, \ref{eq:powerlaw} and \ref{eq:sinhclaw}) with the measured behavior of several different single photon counting detectors (SPCM-AQ4C from PerkinElmer (see Figure~\ref{fig:fitting}a), SPCM-NIR from Excelitas (see Figure~\ref{fig:fitting}c) and $\tau$-SPAD-fast from Laser Components (see Figure~\ref{fig:fitting}e). We show that none of these canonical models are universal and vary between individual detectors as well as makes of detectors. We also report on the presence of several higher order afterpulses. An investigation of the region of reduced detection efficiency called the ``dead-time'' found between the signal pulse and the afterpulse shows yet another characteristic variation between makes of detectors, which is not part of the models used so far.

\section{\label{sec:setup}Experimental Setup}
\begin{figure}[ht]
\centering
\includegraphics[width=0.65\textwidth]{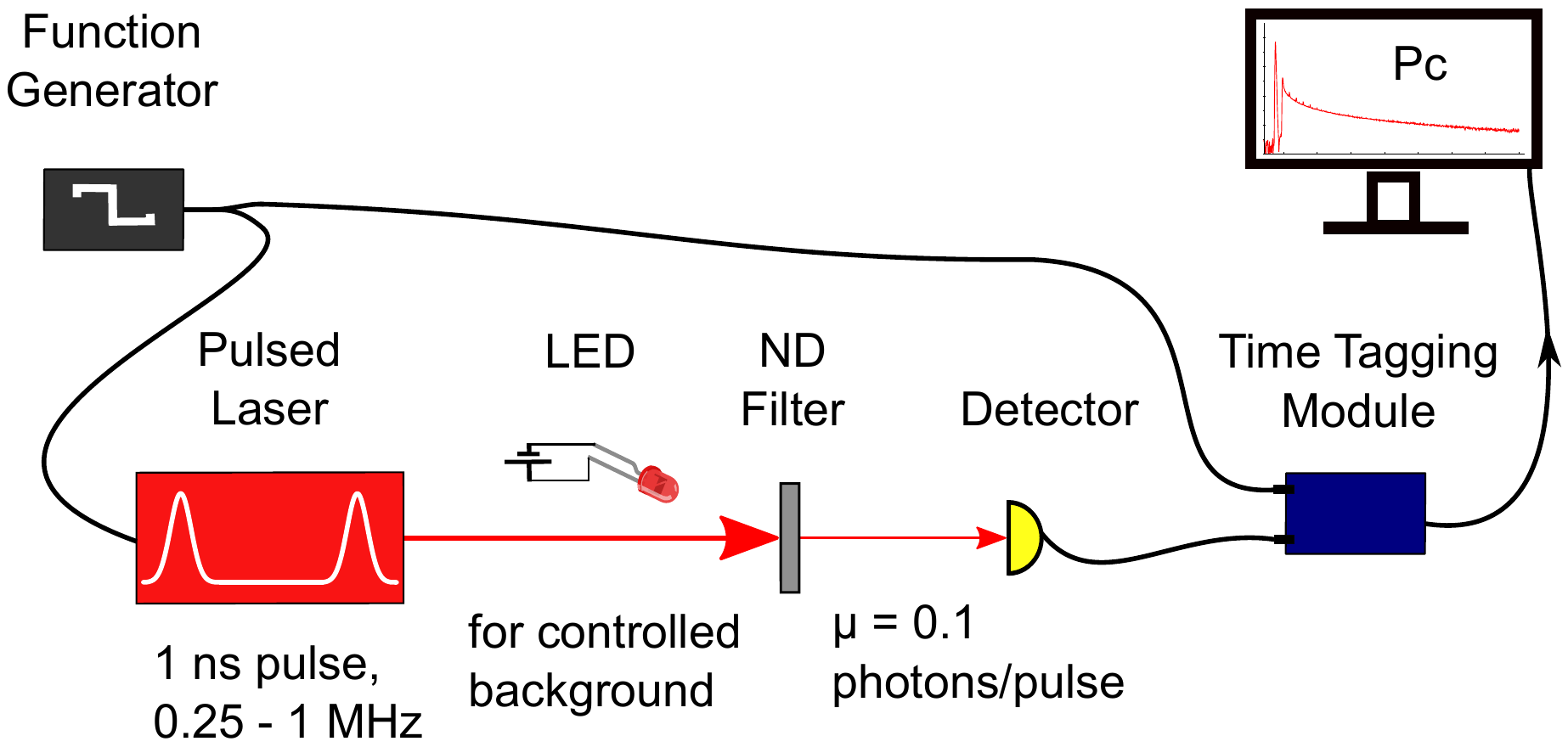}
\caption{ 
A 798\,nm laser is used to generate pulses with a 1\,ns pulse width and various repetition rates between 0.25 to 1\,MHz. We use Neutral Density (ND) filters to attenuate the laser pulse, incident on the active area of the SPADs, until we have $\ll$ 1 detected photon per laser pulse (approximately corresponding to a mean photon number of 0.1 per pulse). We use a time tagging module (TTM8000) with a resolution of 82.3\,ps to record the trigger for the laser pulse (Channel 1) and the photon detection event (Channel 2). This data is stored on a computer and a software computes the temporal cross-correlation histogram ($g^{(2)}$) with 1\,ns bin width. The LED shown was to control the background noise level and was only used to obtain the results shown in Section~\ref{sec:noise}.}
	\label{fig:setup}
\end{figure}

The general scheme of the setup used for the experiments is shown in Figure~\ref{fig:setup}. A function-generator triggers a 798 nm laser with a repetition rate of 0.25 to 1\,MHz. The emitted light pulse of $\approx$ 1\,ns length gets attenuated in a neutral attenuator to $<$0.1\, photons per pulse on average and arrives at the active area of the respective detector in use. The arrival time at the detector and the trigger from the function generator are individually sent to a time-tagging-module (TTM). We then calculated the temporal cross-correlation histogram ($g^{(2)}$) between those two signals using a bin width of 1\,ns. We normalize to the number of detected laser pulses to obtain the pulse probability per bin. To ensure that our results are not skewed by an individual ``faulty'' detector we tested at least 2 detectors of each type (same manufacturer's model number with nearly consecutive serial numbers) all of similar ages. 

We consider the first event arriving at the TTM, after a given laser pulse to be the ''detection pulse'' which stems from the real photon impinging on the active area of the detector. Any detection event after that, but before the next laser pulse, should be due to either afterpulsing or noise. From an end user perspective it is impossible to distinguish a true afterpulse from another count that just happened to occur at a similar time delay. Consequently, we do not differentiate between an afterpulse due to processes in the semiconductor diode from those due to secondary avalanches caused by any other means (such as faults in the quenching circuit). This is typical of most applications and previous studies of these detectors ranging from (quantum) optics experiments to fluorescence microscopy.

We fit each of the standard models (see Equations~\ref{eq:multiexp},~\ref{eq:powerlaw} and Equation~\ref{eq:sinhclaw}) to the tail of the afterpulse as seen in the $g^{(2)}$ obtained for each different SPAD detectors. Figure~\ref{fig:fitting} shows the resultant fits as well as the residuals.
 
All current models of afterpulsing discussed, concern the timing structure of the afterpulse tail. An ideal model of a detector should include its entire behaviour including the duration of the dead time, the jitter and the rising edge of the afterpulse peak. As in most previous previous studies, we then isolate the peak (due to afterpulses) and its tail in time. The tail of the afterpulse of all detectors are shown in Figure~\ref{fig:fitting} after normalization and accidentals correction have been applied. The accidental rate is estimated using a region well after the end of the afterpulse tail and before the next incident photon (this method of correction has been empirically verified in Section~\ref{sec:noise}). We define the total afterpulse probability to be the sum total of the afterpulse probability in each bin of the corrected $g^{(2)}$ histogram for a duration of 900\,ns.~\cite{Note2}. Regardless of the model, in order to obtain a good fit we had to ignore the first two points of the afterpulse tail. 


\section{Results}
\subsection{\label{sec:fitting results}Comparison of afterpulsing models}

{We observed a wide variation between different tested detectors. We tested at least 2 and up to 12 identical detectors (in terms of make, model number and year of purchase)  from each of three manufacturers. We saw a drastic change in the total afterpulse probability between otherwise identical detectors
(see Table~\ref{tab:approb}).}

\begin{figure*}[ht]
	\centering
    \begin{subfigure}[b]{0.45\textwidth}
        \includegraphics[width=\textwidth]{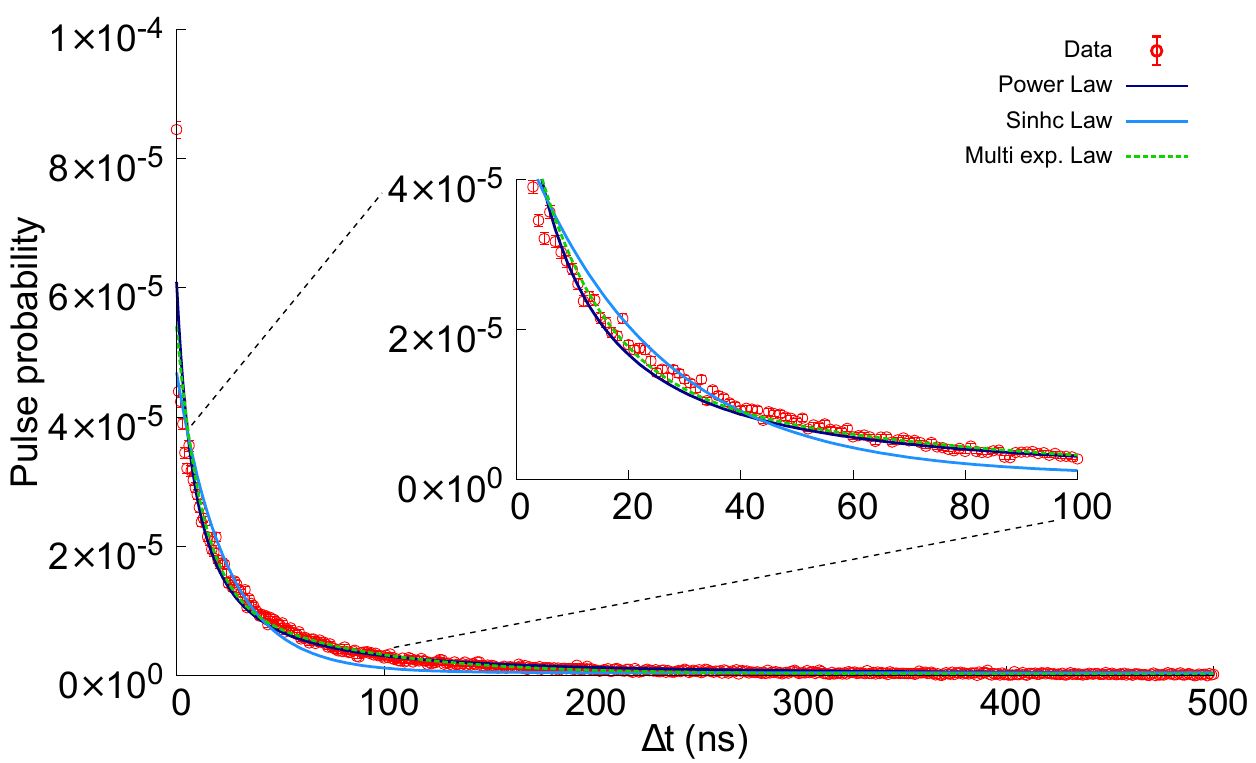}
        \caption{SPCM-AQ4C}
       
    \end{subfigure}
    \begin{subfigure}[b]{0.45\textwidth}
        \includegraphics[width=\textwidth]{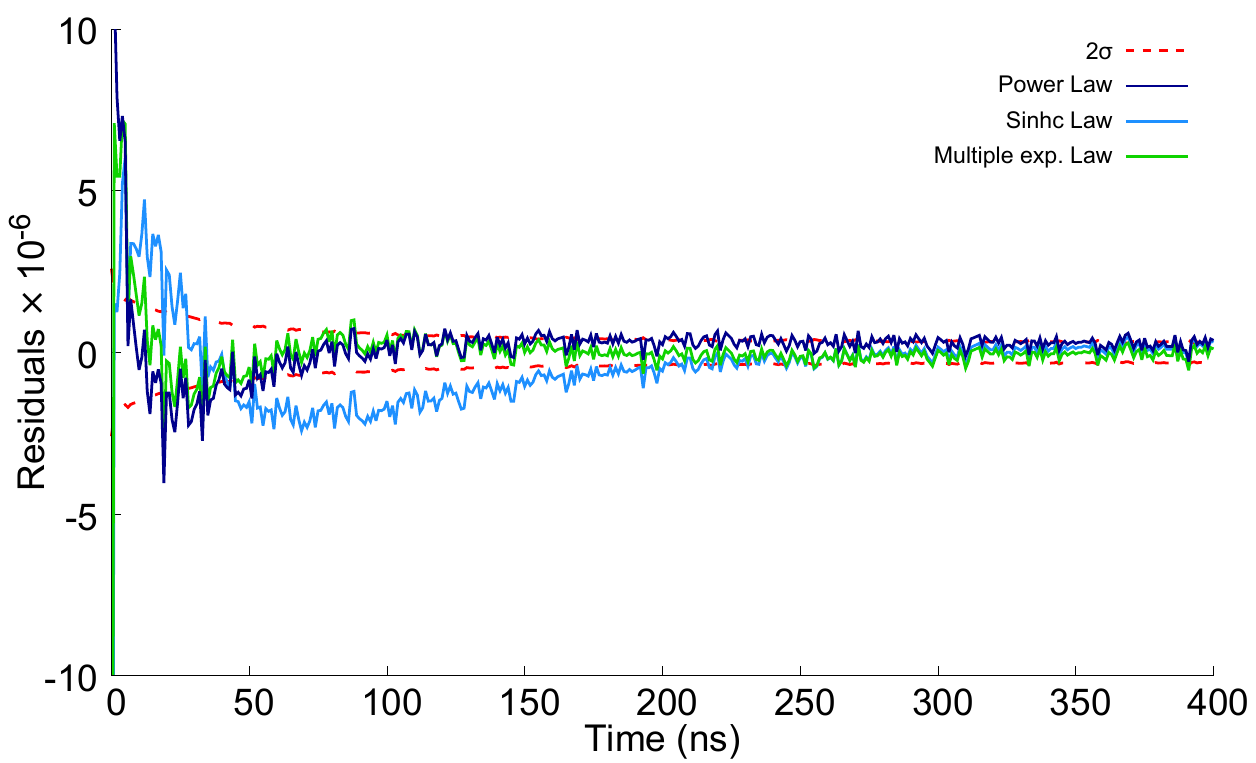}
        \caption{SPCM-AQ4C}
       
    \end{subfigure}
    \begin{subfigure}[b]{0.45\textwidth}
        \includegraphics[width=\textwidth]{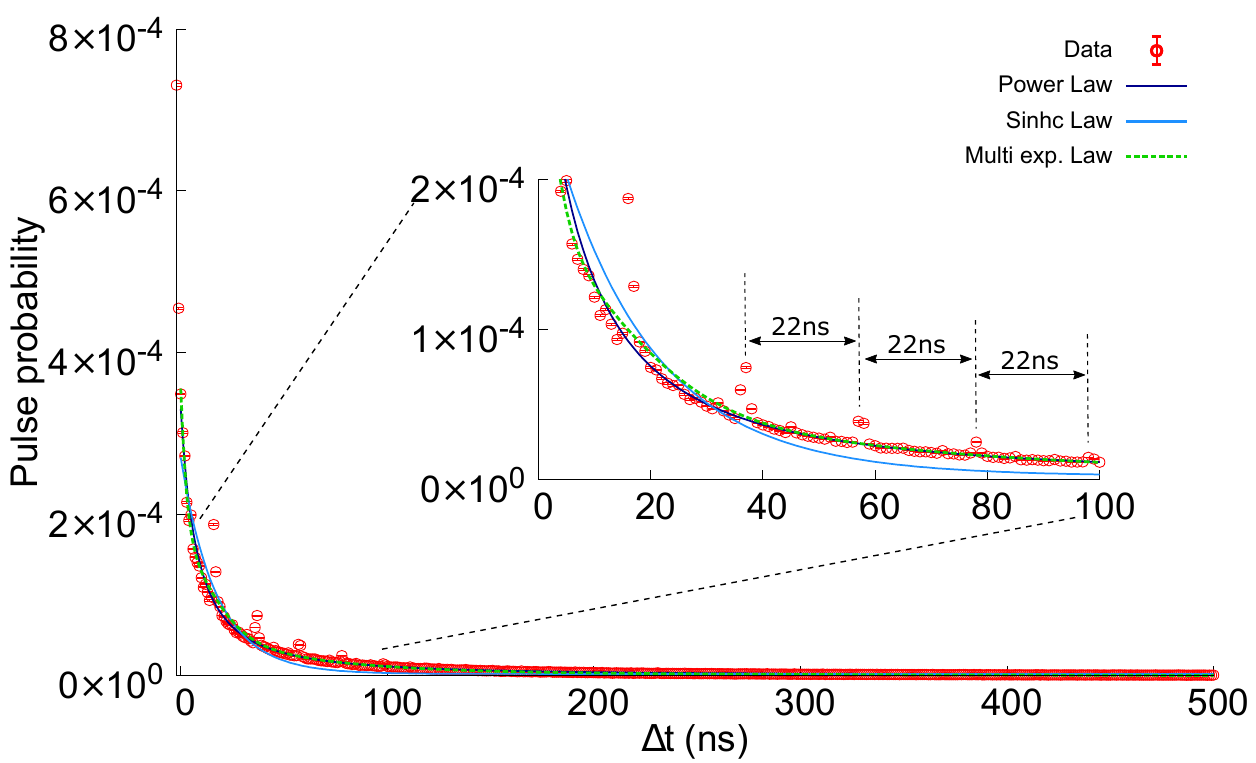}
        \caption{SPCM-NIR}
        
    \end{subfigure}
    \begin{subfigure}[b]{0.45\textwidth}
        \includegraphics[width=\textwidth]{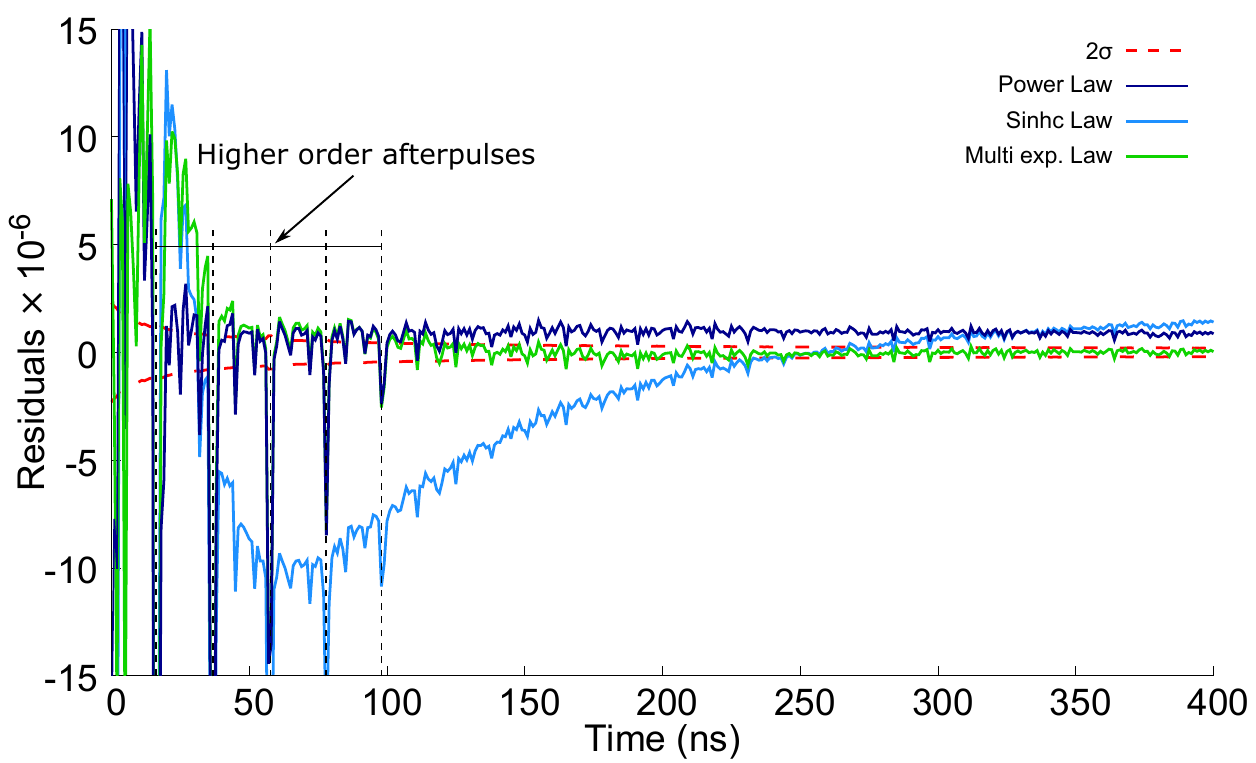}
        \caption{SPCM-NIR}
       
    \end{subfigure}  
    \begin{subfigure}[b]{0.45\textwidth}
        \includegraphics[width=\textwidth]{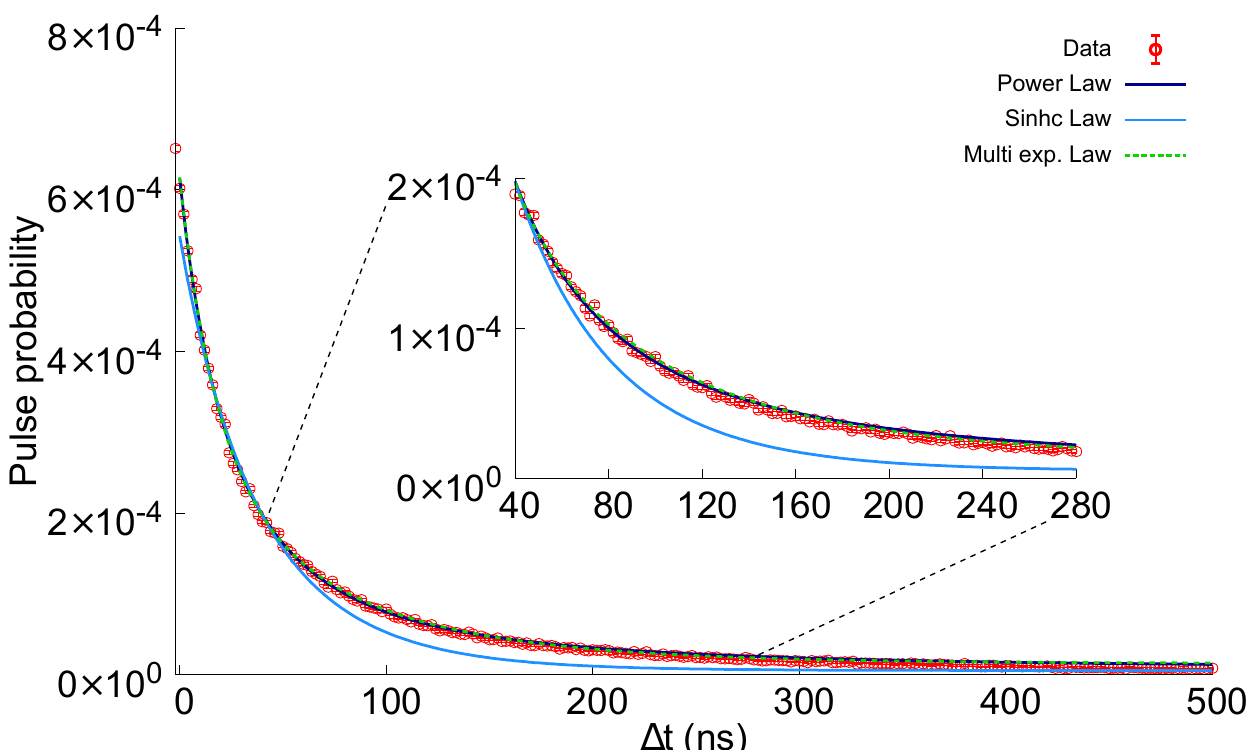}
        \caption{$\tau$-SPAD-fast}
      
   \end{subfigure}
    \begin{subfigure}[b]{0.45\textwidth}
        \includegraphics[width=\textwidth]{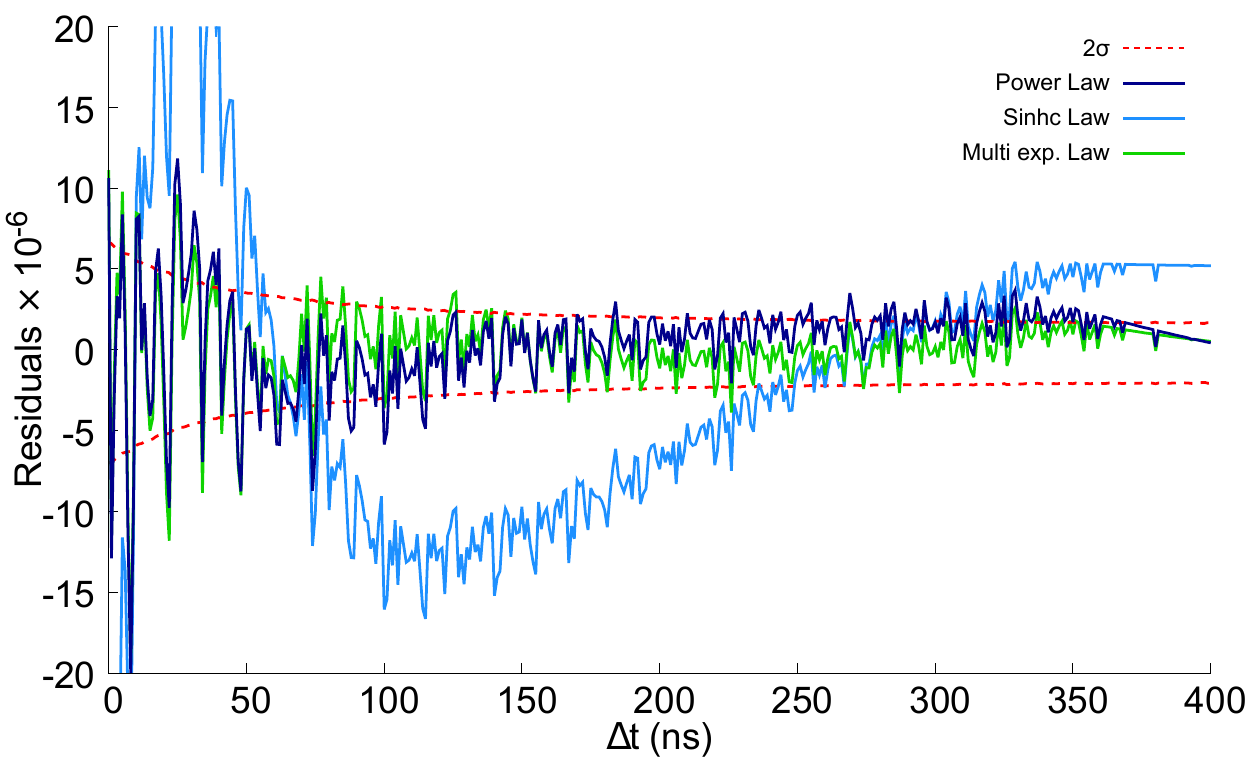}
        \caption{$\tau$-SPAD-fast}
       
    \end{subfigure}
  	\caption{Afterpulse peak's tail fitted with Power law model, Hyperbolic-sinc model and multiple exponential model for SPCM-AQ4C (PerkinElmer), SPCM-NIR (Excelitas) and $\tau$-SPAD-fast (Laser Components) - Comparison of residuals for all detectors and models (Power law model (blue), Hyperbolic sinc model (cyan) and Multiple exponential model (green)) and the red dashed lines are $\pm$ 2 standard deviation limits for statistical fluctuations (2\,$\sigma$). We analysed $> 0.14 \times 10^6, 2.7 \times 10^6$ and $3 \times 10^6$ \emph{afterpulse} events for each SPCM-AQ4C, SPCM-NIR, and $\tau$-SPAD-fast detector.
  	}\label{fig:fitting}
\end{figure*}

On the whole, the power law model had the most consistent good fit and the hyperbolic Sinc model the worst.  Unlike the hyperbolic Sinc  model, the power law model approximates the experimental data more accurately independent of the detector used AQ4C (see Figure~\ref{fig:fitting}a), SPCM-NIR  (see Figure~\ref{fig:fitting}c), $\tau$-SPAD-fast  (see Figure~\ref{fig:fitting}e). Ref.~\cite{Horoshko2015} showed that the hyperbolic sinc model was better than the power law model using id100-MMF50 SPAD module from idQuantique, we were able to duplicate those results only with 6 of the 16 detectors we tested.

Figure~\ref{fig:fitting} shows sample fits of the three models for one detector from each manufacturer.
In this case the power law model is fitting the entire range of the tail better than the Hyperbolic sinc model. The fits can be quantitatively compared using the residuals seen in Figure~\ref{fig:fitting} which are well within the $\pm$ 2 standard deviation range (shown in red) for the power law model but not for the hyperbolic sinc model.

Power law model as a simple analytic model function leads to some good results and proved to be the best model for 4 of the 16 detectors tested.
The power law model was primarily shown to be effective for InGaAS/InP detectors~\cite{Itzler2011,Itzler2012,Restelli2012} and we now show that it is a reasonably good model for some Si detectors.
However, it's estimated parameters gives a rather loose description of the physical origin of afterpulses. A truly good model should be able to predict the behavior of the entire tail region. The power law model only provides a suitable fit if the beginning two points of the afterpulse tail are ignored. 

The multiple exponential model provides a good fit when $k \geqslant 5$ and is slightly better than the power law model for the SPCM-AQ4C detector. It proved to be the best fit for the behaviour of 6 of the 16 tested detectors. It is nearly commensurate with the power law model for the $\tau$-SPAD. For the SPCM-NIR, the exponential model gives a good fit only when we ignore the higher order afterpulse peaks (which are discussed in Section~\ref{sec:higherorder}). We observed that it was possible to get different fits of almost the same quality with different characteristic times ($\tau_k$ also called the de-trapping times) depending on the number of exponentials used. Thus, upon examining the fitting procedure we concur with~\cite{Itzler2012} who state that ``It is evident from this whole fitting procedure that the extracted values for the de-trapping times depend entirely on number of exponentials in the model function and the range of hold-off times used in the data set.''

In case of the hyperbolic sinc model, we consistently obtain unsuitable fits to all but 6 detectors. In figure~\ref{fig:fitting} we can see that the residuals for several fits are much larger than those for other models.   However, for 6 detectors, the  hyperbolic sinc model was found to be the most suitable.


\begin{table}[]
\centering
\label{my-label}
\begin{tabular}{l|l|l|l|l}
\specialrule{.1em}{.05em}{.05em} 
\begin{tabular}[c]{@{}l@{}}\textbf{Make} \\ \textbf{of detector}\end{tabular}   & \begin{tabular}[c]{@{}l@{}}\textbf{Serial}\\ \textbf{number}\end{tabular} & \begin{tabular}[c]{@{}l@{}}\textbf{Factory}\\ \textbf{date}\end{tabular} & \textbf{P(AP)\,(\%)}                                             & \textbf{Best model}                                            \\  \specialrule{.1em}{.05em}{.05em} 
\multirow{24}{*}{\begin{tabular}[c]{@{}l@{}}Excelitas \\ SPCM-NIR\end{tabular}   }                & 29860                                                                     & 27\,Aug.\,15                                                             & \begin{tabular}[c]{@{}l@{}}1.29043\\ \,$\pm$0.00071\end{tabular} & \multirow{3}{*}{\begin{tabular}[c]{@{}l@{}}Multiple\\ exponential\end{tabular}} \\ \cline{2-4}
                                                                                & 29864                                                                     & 27\,Aug.\,15                                                             & \begin{tabular}[c]{@{}l@{}}0.01130\\ \,$\pm$0.00025\end{tabular} &  \\  \cline{2-5}
                                                                                & 32403                                                                     & 26\,Oct.\,16                                                             & \begin{tabular}[c]{@{}l@{}}0.87774\\ \,$\pm$0.00028\end{tabular} & \multirow{8}{*}{Power}                                                          \\ \cline{2-4}
                                                                                & 32404                                                                     & 26\,Oct.\,16                                                             & \begin{tabular}[c]{@{}l@{}}2.01987\\ \,$\pm$0.00039\end{tabular} &                                                           \\ \cline{2-4}
                                                                                & 32405                                                                     & 26\,Oct.\,16                                                             & \begin{tabular}[c]{@{}l@{}}1.38750\\ \,$\pm$0.00033\end{tabular} &                                                           \\ \cline{2-4}
                                                                                & 32406                                                                     & 26\,Oct.\,16                                                             & \begin{tabular}[c]{@{}l@{}}1.56231\\ \,$\pm$0.00035\end{tabular} &                                                           \\ \cline{2-5}
                                                                                & 32424                                                                     & 02\,Nov.\,16                                                             & \begin{tabular}[c]{@{}l@{}}2.91185\\ \,$\pm$0.00048\end{tabular} & \multirow{12}{*}{\begin{tabular}[c]{@{}l@{}}Hyperbolic\\ sinc\end{tabular} }     \\ \cline{2-4}
                                                                                & 32425                                                                     & 02\,Nov.\,16                                                             & \begin{tabular}[c]{@{}l@{}}3.96425\\ \,$\pm$0.00057\end{tabular} &       \\ \cline{2-4}
                                                                                & 32426                                                                     & 02\,Nov.\,16                                                             & \begin{tabular}[c]{@{}l@{}}4.30335\\ \,$\pm$0.00059\end{tabular} &       \\ \cline{2-4}
                                                                                & 32427                                                                     & 02\,Nov.\,16                                                             & \begin{tabular}[c]{@{}l@{}}4.69090\\ \,$\pm$0.00060\end{tabular} &      \\ \cline{2-4}
                                                                                & 32428                                                                     & 02\,Nov.\,16                                                             & \begin{tabular}[c]{@{}l@{}}3.31100\\ \,$\pm$0.00051\end{tabular} &       \\ \cline{2-4}
                                                                                & 32429                                                                     & 02\,Nov.\,16                                                             & \begin{tabular}[c]{@{}l@{}}2.98385\\ \,$\pm$0.00049\end{tabular} &       \\ \specialrule{.1em}{.05em}{.05em} 
\multirow{3}{*}{\begin{tabular}[c]{@{}l@{}}PerkinElmer\\ SPCM-AQ4C\end{tabular} }                & 195 (Ch 1)                                                                       & 25\,Jan.\,07                                                             & \begin{tabular}[c]{@{}l@{}}0.28542\\ \,$\pm$0.00079\end{tabular} & \multirow{3}{*}{\begin{tabular}[c]{@{}l@{}}Multiple\\ exponential\end{tabular}} \\ \cline{2-4}
\begin{tabular}[c]{@{}l@{}}~\end{tabular}                 & 195 (Ch 2)                                                                      & 25\,Jan.\,07                                                             & \begin{tabular}[c]{@{}l@{}}0.28542\\ \,$\pm$0.00079\end{tabular} &  \\ \specialrule{.1em}{.05em}{.05em} 
\multirow{3}{*}{\begin{tabular}[c]{@{}l@{}}Laser \\ Components \\ $\tau$-SPAD-fast\end{tabular}} & 1019917                                                                   & 01\,Nov.\,13                                                             & \begin{tabular}[c]{@{}l@{}}5.10601\\ \,$\pm$0.00310\end{tabular} & \multirow{3}{*}{\begin{tabular}[c]{@{}l@{}}Multiple\\ exponential\end{tabular}} \\ \cline{2-4}
                                                                                & 1019920                                                                   & 01\,Nov.\,13                                                             & \begin{tabular}[c]{@{}l@{}}8.52410\\ \,$\pm$0.00213\end{tabular} &  \\ \specialrule{.1em}{.05em}{.05em} 
\end{tabular}
\caption{{Table comparing various makes of commercially available single photon detectors we tested. The total after pulse probility shows a piece to piece variability as expected. More importantly, the piece to piece variation extends to the best fitting theoretical model. This should not happen for theories based on fundamental properties of semiconductors. Clearly the applicability of all 3 mathematical models discussed here is very limited and ill defined.}}
    \label{tab:ap_all}\label{tab:approb}
\end{table}

{The total probability of obtaining an afterpulse ($P(AP)$) for each detector is shown in Table~\ref{tab:ap_all}. 
Clearly, afterpulse behavior and probabilities vary drastically between brands and between individual specimens of the same brand.
Various detectors of the same make from the same company (even if manufactured 1 week appart) provide evidence in support of different and contradictory mathematical models. As do detectors from different brands. Clearly a universal model to describe afterpulsing behaviour does not exist and every detector needs to be characterized individually.}

\subsection{\label{sec:higherorder} Higher order afterpulses}

\begin{figure}[ht]
\centering
\includegraphics[width=0.65\textwidth]{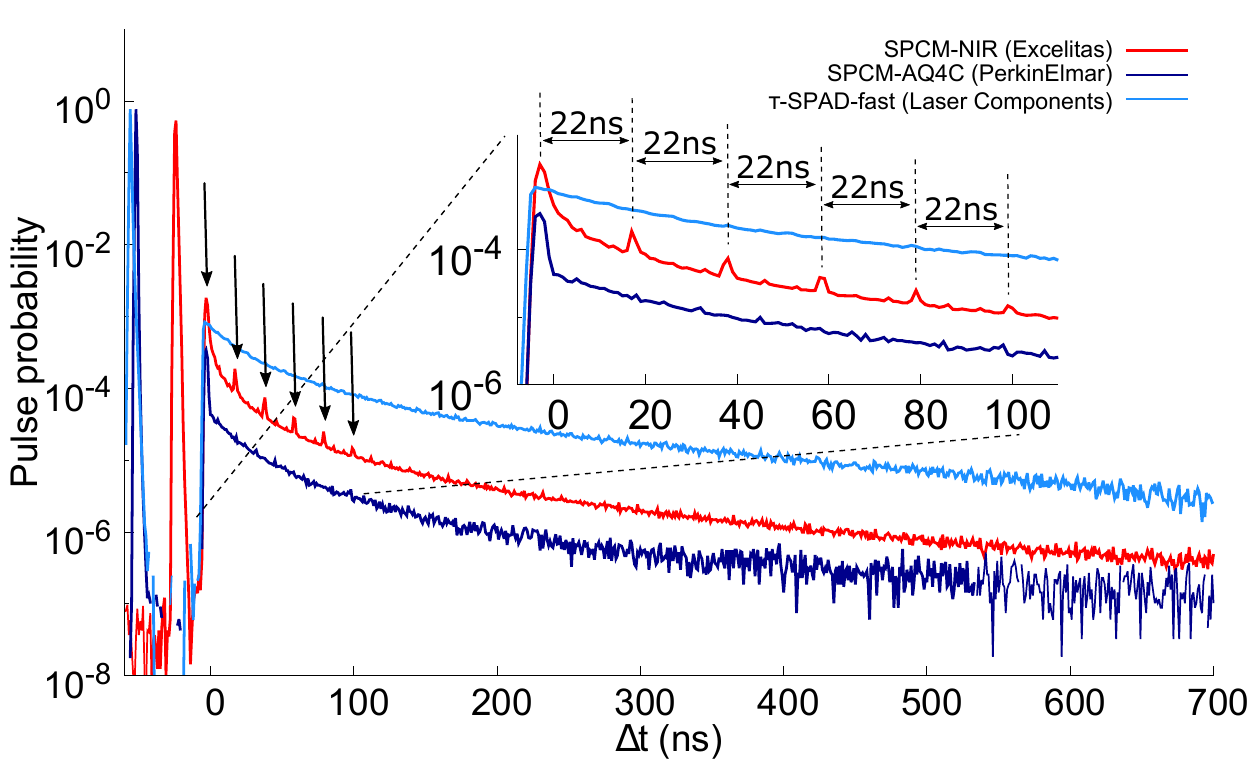}
	\caption{g$^{(2)}$ Histograms for various detectors exhibit distinct afterpulse behavior (each afterpulse peak is marked by an arrow). For example, the $\tau$-SPAD-fast displays an unusually gradual decay while the SPCM-NIR is the only detector make to exhibit higher order afterpulses . The inset shows the higher order afterpulses occurring at intervals equal to the dead time.
}
	    \label{higherorder}
\end{figure}

{Any detection event may cause an afterpulse which, being a detection event, may induce secondary and further afterpulses called ``higher order afterpulses''. Usually afterpulsing is a small effect and the afterpulsing probability distribution function is smeared such that the higher order afterpulses are improbable. However, in the presence of strong enough twilighting~\cite{Polyakov2007}, photon detections and afterpulses accumulate in a narrow peak that appear just after the dead time. We note this behavior in SPCM-NIR and to a lesser extent in SPCM-AQ4C, as can be seen in Figure~\ref{higherorder}. The higher order afterpulses are clearly visible as a series of peaks after the main peak with a period exactly equal to the dead time of the detector. } The presence of such higher order afterpulsing was first speculated in Ref.~\cite{Itzler2012} but, to the best of our knowledge, never reported.

{For the SPCM-NIR, the signal pulse and the first order of afterpulse are 22\,ns apart, which exactly corresponds to the duration of the dead time of this particular detector module. The measured time intervals between all following higher orders of afterpulses (as seen in the inset of the Figure~\ref{higherorder}) have the same time delay of 22\,ns.  We have obtained a similar plot for three detectors with the same model number: in each case, peaks appear separated by the dead time of the particular detector. Such a behavior is clearly undesirable, notably in time-resolved spectroscopy where higher order peaks could be mistaken for, or mask the true signal.}

We verified that the higher order afterpulses seen are not the result of stray light, the shape of the laser pulse, electronic noise, impedance mismatch or optical reflections.
We also processed the data with different bin widths and laser pulse frequencies to ensure that the observed higher order afterpulse peaks are not due to digitization noise.

Further, we note that it is possible to explain the area of the $n^{\mathrm{th}}$ higher order peak based on the probability of the first afterpulse ($P(AP)$) as $P(AP)^{(n+1)} + $ the probability of an afterpulse in the bin just before the  $n^{\mathrm{th}}$ higher order peak $\times$ the number of bins in the peak.
This geometric progression agrees to within 4 to 6\,\% for the $2^{nd}$ to $5^{th}$ order afterpulses.

\subsection{\label{sec:noise} Background and accidentals corrections during the dead time}

\begin{figure*}[t]
	\centering
	
	\begin{subfigure}[t]{0.45\textwidth}
		\includegraphics[width=\textwidth]{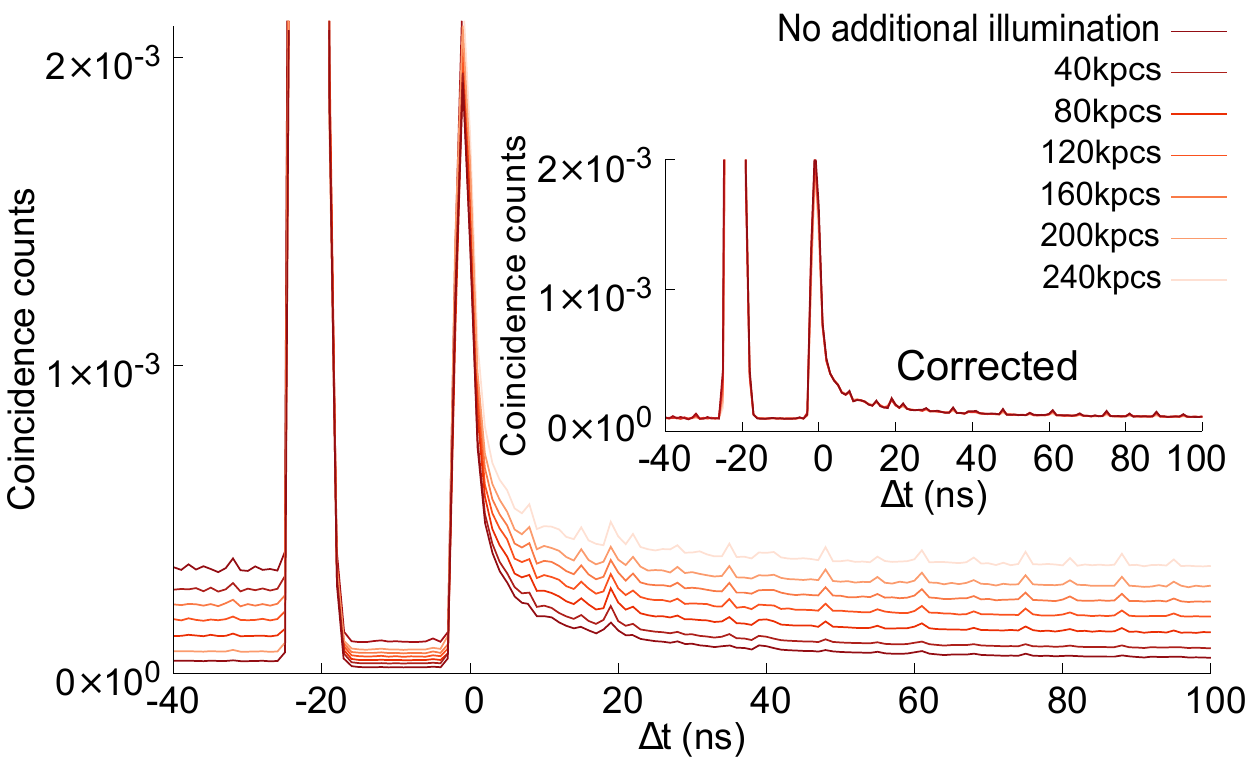}
	\caption{Excelitas SPCM-NIR APD module.}
		\label{bkgrnd_NIR}
	\end{subfigure}
	\begin{subfigure}[t]{0.45\textwidth}
		\includegraphics[width=\textwidth]{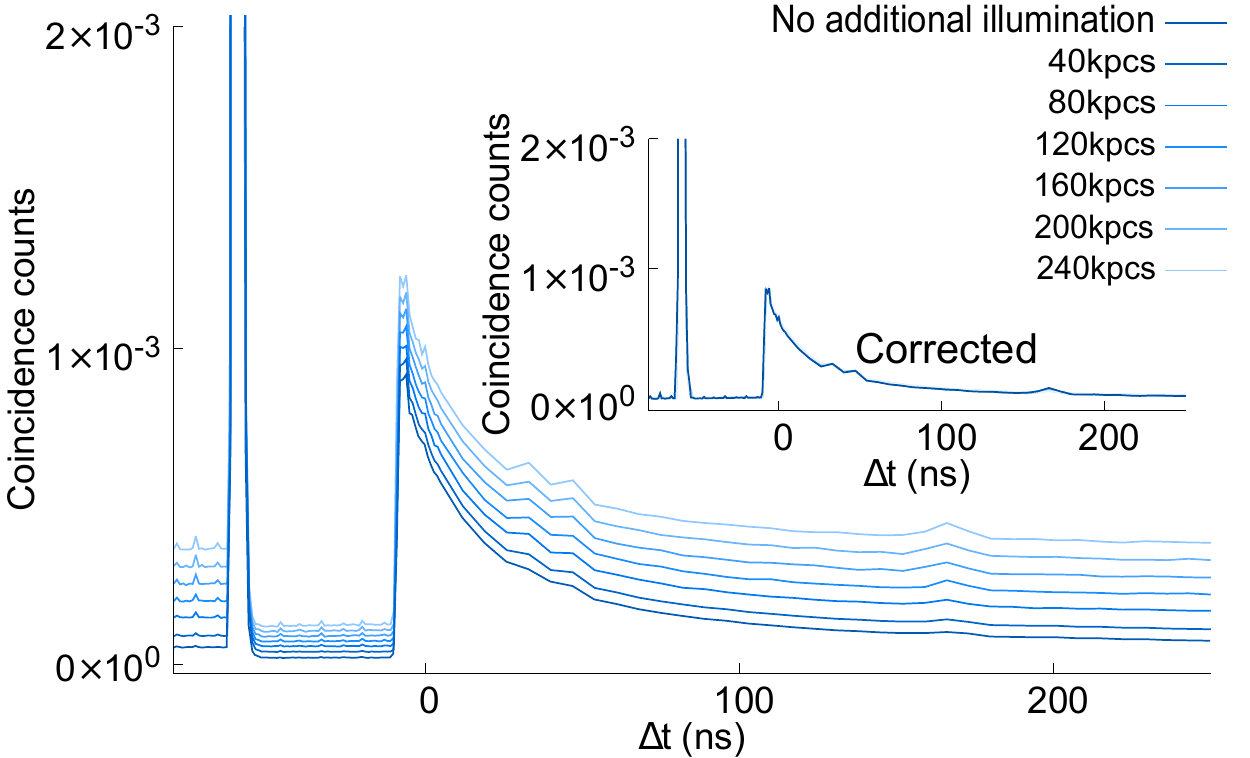}
	\caption{Laser Components $\tau$-SPAD-fast APD module.}
	\label{bkgrnd_tau}
	\end{subfigure}
	
	\caption{g$^{(2)}$ Histograms for different  background count levels, clearly showing the probability of detection events during dead time region during strong background illumination (between the large detection peak and time $\Delta t = 0$). We observe a linear scaling of the accidental coincidences as we vary the background count rate. {\bf Insets}: The g$^{(2)}$ histograms corrected for background counts.}
	\label{bkgrnd}
	
\end{figure*}

In this subsection we discuss the corrections that we can  apply to the cross correlated histograms we used in the previous subsections. 
In typical quantum optics experiments, there is a probability that a coincidence is detected between two different detectors erroneously, we call these coincidences ``accidentals''. Typically, they can be estimated from Poissonian statistics as:
$r_{acc} = r_1 r_2 t_c$, where $r_{acc}$ is the rate of the accidental counts, $r_1$ and $r_2$ are the count rates of the individual detectors and $t_c$ is the coincidence time window used. By introducing a controlled amount of continuous background illumination we experimentally verify this standard practice. In the duration well after a detection event, this provides a very good estimate (with a maximum variation of $\ll$ 3\,\%) of the behavior (see Figures~\ref{bkgrnd_NIR}~and~\ref{bkgrnd_tau}).

An interesting consequence of using the cross-correlation technique described above is the non-zero probability of a count during the ``dead-time''. We attribute these detection events to the probability that the detector did not click because of the laser pulse but did click within the dead-time region.

When used for some tasks (like quantum communication), the end user may not be able to correct for the accidentals obtained while computing the cross correlation histogram. In these cases it is important, for a complete model of the detector's behavior, that these accidentals during the dead-time be correctly accounted for.

We use a continuous wave battery powered LED as a steady and controllable source of background illumination in addition to the attenuated laser pulses (see Figure~\ref{fig:setup}). 
 We then measured the $g^{(2)}$ histograms for different background count rates.
 Figure~\ref{bkgrnd_NIR} shows these histograms for the SPCM-NIR while Figure~\ref{bkgrnd_tau} shows the same for the $\tau$-SPAD-fast modules~\cite{Note1}.
We show that the increase in click probability during the dead-time scales linearly with the amount of additional background illumination (see Figures~\ref{bkgrnd_NIR}~and~\ref{bkgrnd_tau}).

 In the regions before the detection event, well after the detection event and even during the afterpulse, the measured accidental corrections agreed with the value calculated assuming Poissonian statistics.

\section{Conclusion}
We have clearly demonstrated that different standard models are required to appropriately describe the distribution of electrical signals generated by different detectors. This explains the conflicting nature of several previous studies; for example Refs.~\cite{Itzler2012,Itzler2011,Restelli2012} show strong evidence for the power law model while Refs.~\cite{Giudice2003,Cova1991} show equally compelling evidence for the multiple exponential model and Ref.~\cite{Horoshko2015} provides evidence for the hyperbolic sinc model. {Recently~\cite{Wang2016} showed that the afterpulse probability is dependent on past events --- a property not considered in the exponential, power, or hyperbolic sinc models.} By comparing previously reported results to our own, we realize that there is a large variation between the different commonly used manufactures/makes of detectors and between individual detectors. The suppression of afterpulses by different quenching methods~\cite{Eisaman2013} leads us to believe that the afterpulsing behavior is more dependent on the electronic quenching circuit used rather than the properties of the semiconductor (such as the presence and distribution of discrete/continuous/quasi-continuous deep levels). Our data clearly proves that none of the current theoretical models are universal which makes it hard to draw conclusions about the underlying mechanism based on fundamental semiconductor physics.

{As seen in Table~\ref{tab:ap_all} supposedly identical detectors manufactured within a week of each other, provide evidence to support one model over the other. Several previous studies have shown that one or the other mathematical model fits the afterpulsing behaviour of a few individual detectors. Further many of these studies use such evidence to confirm or disprove hypotheses/assumptions about semiconductor physics. Clearly, with such extreme individual variation no conclusion about the physics behind these mathematical models can be drawn from any  similar test. Especially tests with a small number of individual detectors or with detectors of only one kind.}

We also report on the presence of higher order afterpulses in one of the tested detector models. To make this possible, unlike in several previous measurements~\cite{Horoshko2015,Humer2015,Stip2013} who used an auto-correlation signal, we use a cross-correlation histogram between the detector and the trigger. This allows us to look for both higher order afterpulses as well as the behavior during the dead time. These higher order afterpulses can cause large errors in measurements of the arrival times of photons and must be carefully accounted for. We were able to exclude frequency dependence in our detectors since we repeated all the experiments described above for several different repetition rates (all far from detector saturation) of the laser pulses ranging from 10\,kHz to 1.2\,MHz and found no significant variation.

{Further, the total afterpulsing probability for several individual detectors all of the same manufacturer and part number varied drastically (in some cases by a factor of 460!) implies that every individual detector needs to be calibrated for all applications that need to accurately measure count rates or arrival times of photons.}

For many years, afterpulsing has been extensively studied from a semiconductor physics based perspective, where it is important to understanding how trapped charges/energy levels decay. However, to correctly study this, one must separate the effects due to the behaviour of the diode or seimiconductor junction from the effects due to the electronics and quenching circuits. {For example, based on modeling the semiconductor junction, a longer dead time is thought to lead to a lower afterpulsing probability~\cite{Humer2015}. However, as seen in Table~\ref{tab:approb} and Figure~\ref{higherorder} both the SPCM-AQ4C and the $\tau$-SPAD-fast have a significantly longer dead time than the SPCM-NIR but the former has a lower afterpulse  probability while the latter has a larger one. In this case it is impossible to draw a conclusion about the relationship between dead time and afterpulse probability without carefully considering the quenching circuits used. This type of analysis requires proprietary and confidential information about the diode and circuits used by the manufacturer; which is inaccessible to a typical end-user. 
Instead of drawing potentially erroneous conclusions about semiconductor behavior based on which model had a better fit, we focus on an application oriented perspective. For most practical purposes, it is sufficient to understand the statistical nature of the afterpulses rather than their causal mechanisms.}

We also would like to mention, that we did not consider a possible aging effect of the detectors. All measurements were performed in the time span of only a few months. Commercial products, may have to be characterized repeatedly during their long operational lifetimes. The study of any aging effects on the statistical behaviour of SPAD's in general is an interesting avenue for further exploration. 

{Most applications of SPADs are hindered by afterpulsing, in many cases these ill effects can be corrected for if each individual detector is properly characterized. The individual characterization is necessary due to the large variation in both the total afterpulse probability and the mathematical form of the probability distribution between detectors of the same model number, age and manufacturer under nearly identical laboratory test conditions. It is possible, although inadvisable, to ignore all clicks for several hundred ns after any detection event. This effectively increases the dead time and avoids the bulk of afterpulses but this severely limits the maximum count rates and detection efficiency due to saturation-like effects. To correctly account for the ill effects of afterpulses it is sufficient to characterise each individual detector prior to/during use, using the method described in this work.} 
 
{From a quantum communication perspective, our method of characterization of  every individual detector can be included into the overall device dependent security analysis~\cite{gottesman2004security,gottesman2002security}. These characterizations can also be used to improve the accuracy of results in quantum meterology.}

\section*{Acknowledgements}
The Authors thank FFG-ALR (contract Nr. 844360), ESA (contract Nr. 4000112591/14/NL/US) and {MoSES 533-19-14-0002},  the Austrian Academy of Sciences as well as Austrian Science Fund (FWF): SFB F40 (FOQUS) for their financial support. We would like to thank Dmitri Horoshko, Vyacheslav Chizhevsky and Sergei Kilin for discussing their methods and sharing the results of their paper~\cite{Horoshko2015}.

\end{document}